## Title:

Multimodal Learning and Cognitive Processes in Radiology: MedGaze for Chest X-ray Scanpath Prediction

## Author list:

**Akash Awasthi**
PhD student
Department of Electrical and Computer Engineering
University of Houston
akashcseklu123@gmail.com

**Ngan Le, Ph.D.**
Assistant Professor
Department of Computer Science & Computer Engineering
University of Arkansas
thile@uark.edu

**Zhigang Deng, Ph.D.**
Moores Professor of Computer Science
Department of Computer Science,
University of Houston, Houston, TX
zdeng4@central.uh.edu

**Rishi Agrawal, MD**
Associate Professor
Department of Thoracic Imaging, Division of Diagnostic Imaging, The University of Texas MD Anderson Cancer Center, Houston, TX
RAgrawal1@mdanderson.org

**Carol C. Wu, MD**
Professor
Department of Thoracic Imaging,
Division of Diagnostic Imaging, The University of Texas MD Anderson Cancer Center, Houston
CCWu1@mdanderson.org

**Hien Van Nguyen, Ph.D.**
Associate Professor
Department of Electrical and Computer Engineering
University of Houston

hvnguy35@central.uh.edu

## Corresponding author information:

**Akash Awasthi**
PhD student
Department of Electrical and Computer Engineering
University of Houston
Room no- N368, Cullen College of Engineering Building-1
4222 Martin Luther King Blvd, Houston, TX 77204

akashcseklu123@gmail.com

# Abstract

**Background:** Predicting human gaze behavior within computer vision is integral for developing interactive systems that can anticipate user attention and address fundamental questions in cognitive science. While methodologies exist for modeling gaze behavior on natural images, scanpath prediction from radiographic images remains unexplored.

**Purpose:** To develop an AI system that can model the cognitive processes of the radiologist and predict the scanpaths on the CXR images.

**Materials and Methods:** This retrospective study utilized publicly available datasets: REFLACX[1], comprising eye-tracking data from multiple radiologists, and EGD-CXR[2], which includes data from a single radiologist. MedGaze employs a two-stage training process using Large-Multimodal models to generate human-like scanpaths and fixations duration using the radiology reports and CXR images. Evaluation includes metrics such as IoU score, CC score, and Multimatch score, comparing MedGaze against a state-of-the-art method. Human evaluation assessed similarity to human-generated patterns and coverage of the region of interest.

**Results:** MedGaze outperformed the state of the art on both EGD-CXR and REFLACX datasets. On EGD-CXR, MedGaze achieved IoU, CC (Correlation Coefficient), and mean Multimatch score (mMM) of 0.41 [95% CI 0.40,0.42 ] vs 0.27 [95% CI 0.26,0.28 ], 0.50 [95% CI 0.48,0.51 ] vs 0.37 [95% CI 0.36,0.41 ], and 0.80 [95% CI 0.79,0.81 ] vs 0.71 [95% CI 0.70,0.71 ] compared to the state of the art. On REFLACX, MedGaze scored 0.45 [95% CI 0.44,0.46 ] vs 0.30 [95% CI 0.29,0.30 ], 0.53 [95% CI 0.50,0.55 ] vs 0.40 [95% CI 0.38,0.42 ], and 0.84 [95% CI 0.83,0.85 ] vs 0.76 [95% CI 0.75,0.77 ]. MedGaze also demonstrated its ability to assess case difficulty through fixation duration, showing a significant Spearman rank correlation of 0.65 ($p=0.00$) with true case difficulty ranks on EGD-CXR. In human evaluation, 13 out of 20 MedGaze-predicted scanpath videos resembled human-generated patterns, and 18 out of 20 achieved a comprehensive score of 4 (60-80% region coverage). Additionally, MedGaze-predicted scanpaths showed minimal redundancy (redundancy score = 1) compared to human-generated ones (9 out of 20 vs 5 out of 20).

**Conclusion:** Modeling scanpaths on radiology images is crucial for understanding and anticipating radiologist's eye movements, enhancing training standardization, and improving diagnostic accuracy.

# Introduction

The modeling of human gaze behavior is a critical problem in computer vision, with significant implications for designing interactive systems that can anticipate a user's attention. In medical imaging, particularly with chest X-rays (CXR), predicting scanpaths is essential for enhancing diagnostic accuracy and efficiency. By analyzing how expert radiologists navigate these images, we can develop advanced training programs to help novices adopt effective viewing strategies, thereby reducing errors and improving their diagnostic skills.

Predicting human scanpaths on medical images presents unique challenges compared to natural images due to the presence of abnormal regions with varying shapes, sizes, and contrasts [6]. Previous research has focused on predicting scanpaths in natural images by targeting specific objects or goals[3,4,5]. Our study introduces MedGaze, a novel system tailored to model scanpaths aligned with radiology reports containing multiple abnormalities. MedGaze predicts fixation points and durations crucial for identifying abnormalities, aiming to enhance human-AI collaboration and refine training modules for novice radiologists.

As shown in Figure 1a, our methodology involves two-stage training: Vision to Radiology Report Learning (VR2) and Vision Language Cognitive Learning (VLC), utilizing large publicly available datasets. Given the limited availability of eye gaze tracking data [1,2], we leverage the MIMIC dataset[7,8] for representation learning to extract medically relevant multimodal features, which are then used to model eye gaze movements. Our model employs Large Multimodal Models (LMMs) to extract text-enriched multimodal embeddings. Unlike previous computer vision efforts that focus on predicting scanpaths based on specific objects or categories, our approach addresses a broader context of modeling scanpath sequences for searching multiple abnormalities in CXR images. Specifically, our method scales up the prediction by an order of magnitude compared to existing state-of-the-art methods.

To validate our approach, we compare its performance to current state-of-the-art methods in computer vision for predicting scanpaths on natural images, using statistical metrics. Additionally, we assess our model's ability to generalize across different radiologists. An expert thoracic radiologist provides ratings based on the comprehensiveness and redundancy of predicted scanpaths to evaluate their clinical relevance.

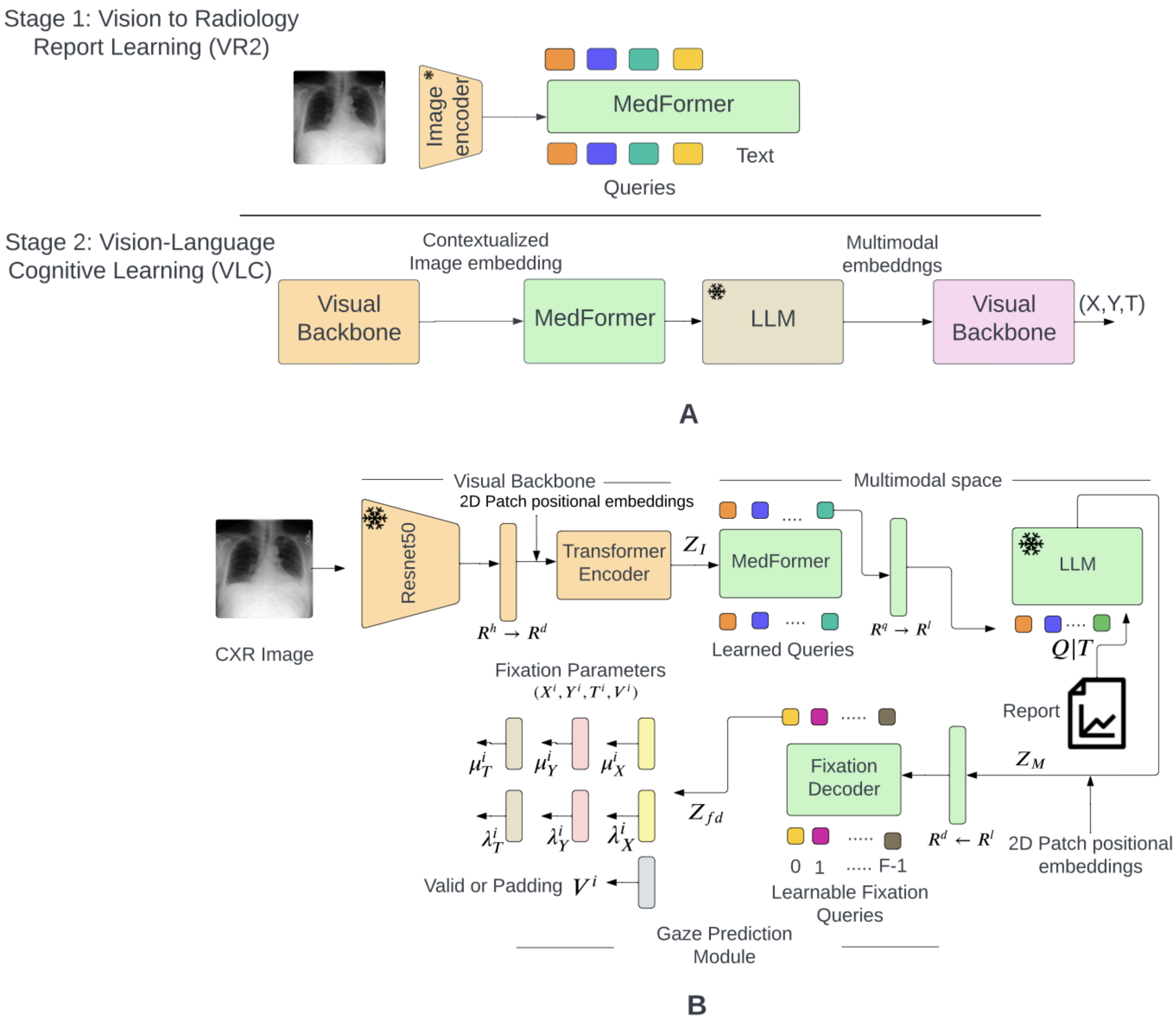

**Figure 1: Representation of the proposed Methodology. Subfigure A illustrates the two-stage training strategy: Vision to Radiology Report Learning (VR2) followed by Vision-Language Cognition Learning (VLC). This approach is designed to extract medically relevant multimodal features and model the cognitive processes of radiologists during CXR image diagnosis. Subfigure B depicts the architecture of the proposed model MedGaze, emphasizing the integration of the Vision-Language Cognitive learning (VLC) stage**

## Materials and Methods:

Figure 1a outlines our two-stage training approach, Vision to Radiology Report Learning (VR2) and Vision-Language Cognition Learning (VLC), aimed at extracting the text-enriched multimodal embeddings to model cognitive processes in CXR diagnosis. Figure 1b expands on our architectural framework, which comprises three pivotal components: the Visual Backbone, the Multimodal Space, and the Gaze Prediction Module.

**Visual Backbone:** The Visual Backbone is essential for extracting contextualized visual embeddings. It includes ResNet-50[9] as a frozen feature extractor that extracts visual features from images. Following this, 6 standard transformer encoder [20] layers are incorporated to generate a contextualized feature embedding, denoted as $Z_I$. Additionally, we employ 2D sinusoidal positional embeddings to denote the location of each patch [10]. Our ablation study experiments reveal that substituting the ResNet-50 (feature extractor) and transformer encoder block with a CLIP-based vision [11]transformer results in increased training duration and computational costs, as well as relatively inferior performance. We include the ablation study results in the supplementary material.

**MedFormer:** During the initial training phase of the VR2, we propose a transformer-based module pre-trained specifically on the MIMIC data, called MedFormer. This module aims to bridge the gap between the frozen image encoder and the large language model, facilitating the extraction of a fixed number of image features irrespective of the input image resolution. It consists of two transformer submodules: one called an image transformer which interacts with the frozen image encoder, and the other one called a text transformer which can function both as a text encoder and decoder. MedFormer filters out unnecessary visual details, providing focused and refined visual context. This reduces the LLM's burden of aligning visual and language data from scratch, making the training process more efficient.

**Large Language Model:** This component serves as the cornerstone of our architecture, tasked with modeling the complex interplay between refined contextualized image embeddings and text embeddings. Consequently, it equips the gaze prediction module with robust multimodal embeddings enriched by textual context. By employing the frozen decoder-based LLM known as OPT[12], we integrate MedGaze's output with text embeddings and input this concatenated representation to the LLM.

**Multimodal Space:** In contrast to the previous Gazeformer [3] model, which employed simple linear projections to create image-text joint embeddings in the visual-semantic space, our experiments demonstrate that this approach falls short for detailed radiology reports. Therefore, we propose to connect MedGaze with a large language model to capture the complex interplay between image and text embeddings. Radiology reports are extensive, detailing numerous diseases or abnormalities that radiologists look for. Thus, simplistic modeling within the visual-semantic space may prove inadequate. The radiology reports show long dependencies since they begin searching for various diseases from the start of the image. Consequently, the sequence in which diseases are identified may involve complex cognitive processes. For example, detecting a patchy opacity of various sizes and shapes could lead to the diagnosis of pneumonia or edema. Our ablation experiments found that the optimal configuration for multimodal space requires integrating both MedGaze and the LLM.

**Gaze Prediction Module**: This module is responsible for predicting both fixation coordinates and fixation duration, and it consists of a fixation decoder and a scanpath prediction network. Specifically, the fixation decoder adopts a transformer decoder-like architecture[20], processing F fixation queries. These learnable fixation queries, which are randomly initialized, encode information about the fixation timestep. The maximum length of the fixation sequence is denoted as F. If the output fixation length is shorter than the maximum sequence length, padding is used to adjust the length to F. We have used 6 standard Transformer decoder layers in the Fixation decoder block. The latent fixation embeddings interact through self-attention and engage with the multimodal embedding (M) via encoder-decoder attention. Furthermore, fixed 2D position encoding is added to the multimodal embedding to provide positional information about the patches.

In the fixation prediction module, fixation coordinates are directly regressed from the output of the fixation decoder $Z_{fd}$, which has a size of $batch\ size \times$ F $\times\ model\ dimensions$, with F indicating the timestep information. Radiologists exhibit variability in gaze sequence patterns, reflecting individual approaches to diagnosing diseases from CXR images, leading to inter-subject variability in fixation patterns. To ensure the model's generalizability across multiple radiologists and avoid learning spurious correlations, fixation coordinates, and durations are modeled using a Gaussian distribution. This involves regressing the mean and log-variance of the 2D coordinates and fixation duration using six distinct MLP layers, employing the reparametrization trick [13]to ensure a fully differentiable network. Padding is employed for fixation sequences shorter than the maximum length set (F), and a separate MLP classifier with a softmax classifier is utilized to predict whether a specific step in the F slices of the multimodal embedding is a valid fixation or a padding token. During inference, (X, Y, T, V) are predicted, where X, Y represent the fixation coordinates, T represents the fixation duration and V represents the probability of this fixation quad being a valid fixation or a padding token. Sequence termination occurs when V>0.5, signaling the start of the padding tokens.

**Training Procedure:** In the initial phase (VR2), we train the MedGaze on the MIMIC data to acquire text-informed vision representation. During this stage, the Qformer[14] is connected with the frozen image encoder to facilitate training using techniques such as Image-Text matching loss [15], Image-Text contrastive loss [14], and Image-Text grounding loss [16]. Moving to the second training stage (VLC), depicted in Figure 1 a, we integrate the MedGaze with the visual backbone (consisting of a frozen image encoder and a transformer encoder ) and the frozen LLM to execute the Vision-Language Cognitive Learning. In this phase, the total loss ($L_t$) is calculated by summing the spatio-temporal loss and the cross-entropy loss for token classification across N samples in the minibatch, as described in Equation 1.

$$L_t = \frac{1}{N}\sum_{k=1}^{N}(L_{spa}^{k} + L_{val}^{k}) \quad (1)$$

Where $L_{spa}^{k} = \frac{1}{l^k}\sum_{i=0}^{l^k-1}(|x_i^k - \hat{x}_i^k| + |y_i^k - \hat{y}_i^k| + |t_i^k - \hat{t}_i^k|)$

$$L_{val}^{k} = -\frac{1}{L}\sum_{i=1}^{L-1}(v_i^k \, log \, \hat{v}_i^k + (1 - v_i^k) log(1 - \hat{v}_i^k))$$

Here, $L_t$ represents the total loss, $L_{spa}$ is the spatio-temporal loss, which is an L1 loss between the predicted and ground truth fixation sequences, including duration. The predicted scanpath, denoted as $s^k = \{(x_i^k, y_i^k, t_i^k)\}_{i=0}^{L-1}$ has a maximum length L, while $l_l^k$ is the length of the ground truth scanpath $\hat{s}^k = \{(x_i^k, y_i^k, t_i^k)\}_{i=0}^{l^k-1}$ . $L_{val}^k$ signifies the validity prediction loss, calculated as the negative log-likelihood loss for validity prediction for each token.

| Dataset | Total samples | Train samples | Test samples |
|---|---|---|---|
| REFLACX | 2507 | 1800 | 707 |
| EGD-CXR | 1072 | 800 | 271 |
| REFLACX + EGD-CXR | 3578 | 2506 | 1078 |

**Table 1:** Representing the number of train/test samples for each dataset

For the VLC training phase, we adopted a batch size of 32 and implemented Disjoint Optimization \cite{mondal2023gazeformer} with the Adam optimizer \cite{kingma2014adam}. This optimization technique employs variable learning rates for different network parameter groups. MedGaze underwent training for 200 epochs to achieve optimal performance.

**Datasets:** In this study, we utilized two datasets: EGD-CXR[2] and REFLACX [1]. These datasets consist of CXR images with synchronized eye-tracking and transcription pairs, annotated by different radiologists. We utilized both datasets to assess the generalization capability of our proposed system. Additionally, we merged both datasets to create a larger dataset, enabling us to evaluate the system's performance comprehensively. Table 1 presents details about the training and testing samples utilized across different datasets. The key hyperparameter we considered was the maximum fixation length, set to 50. This choice was made based on the observation that most cases had a total of 50 scanpaths, indicating that doctors typically concluded their diagnosis within this range. This length is an order of magnitude larger than that of state-of-the-art gaze modeling in natural images [3]. In the supplementary material, we include the distribution plot showing the most common fixation sequence lengths.

**Statistical Metrics:** We assessed our model using two categories of metrics: fixation heatmap-based and scanpath similarity-based evaluations. For fixation heatmaps, we employ Intersection over Union (IoU) and Correlation Coefficient (CC)[17]. IoU quantifies the percentage overlap between the target and prediction masks, while CC gauges the correlation between normalized predicted and human fixation maps. Regarding scanpath similarity, we utilize the mean Multimatch Match Score (MM)[18,19], which aggregates scores for shape, direction, length, position, and duration. Additionally, we present the mD-MM (mean Duration Multimatch score), representing the duration aspect of the MM score and indicating the accuracy of fixation duration predictions. We provide 95% Confidence Intervals derived from the bootstrapped method to ensure the robustness of our findings. For the analysis of case complexity, we compute the Pearson Correlation coefficient for true and predicted total fixation durations, and the Spearman rank correlation coefficient for case difficulty ranks. All statistical calculations were performed using the Scikit-learn package (version 1.2.1) in Python v3.8.

| Method | Train Dataset | Test Dataset | mIoU | mCC | mMM | mD-MM |
|---|---|---|---|---|---|---|
| Gazeformer [3] | EGD-CXR | EGD-CXR | 0.27 (0.26, 0.28) | 0.37 (0.36, 0.41) | 0.71 (0.70, 0.72) | 0.06 (0.04, 0.08) |
| | REFLACX | REFLACX | 0.30 (0.29, 0.30) | 0.40 (0.38, 0.42) | 0.76 (0.75, 0.77) | 0.29 (0.27, 0.33) |
| MedGaze (Ours) | EGD-CXR | EGD-CXR | 0.41 (0.40, 0.421) | 0.50 (0.48, 0.5) | 0.80 (0.79, 0.81) | 0.50 (0.46, 0.52) |
| | REFLACX | REFLACX | 0.45 (0.44, 0.46) | 0.53 (0.50, 0.55) | 0.84 (0.83, 0.85) | 0.66 (0.65, 0.68) |
| Gazeformer [3] | EGD-CXR | REFLACX | 0.26 (0.25, | 0.33 (0.31, | 0.69 (0.68, | 0.07 (0.05, 0.08) |
| | | | 0.27) | 0.34) | 0.70) | |
| | REFLACX | EGD-CXR | 0.28 (0.27, 0.29) | 0.38 (0.36, 0.41) | 0.72 | 0.19 |
| MedGaze (Ours) | EGD-CXR | REFLACX | 0.39 (0.38, 0.40) | 0.42 (0.40, 0.43) | 0.78 (0.77, 0.79) | 0.49 (0.46, 0.52) |
| | REFLACX | EGD-CXR | 0.41 (0.40, 0.43) | 0.50 (0.47, 0.51) | 0.81 | 0.63 |
| Gazeformer [3] | EGD-CXR + REFLACX | EGD-CXR + REFLACX | 0.30 (0.29, 0.31) | 0.42 (0.40, 0.43) | 0.78 (0.77, 0.79) | 0.43 (0.41, 0.45) |
| MedGaze (Ours) | EGD-CXR + REFLACX | EGD-CXR + REFLACX | 0.41 (0.40, 0.42) | 0.49 (0.48, 0.51) | 0.85 (0.84, 0.86) | 0.73 (0.72, 0.74 |

**Table 2: Performance Comparison of MedGaze and Gazeformer on EGD-CXR (single experienced radiologist data) and REFLACX (multiple radiologists data). Values in the bracket represent the 95% Confidence Interval calculated using the bootstrapped method.**

## Results:

Our results section is structured into three distinct parts. Comparison with the state-of-the-art, prediction visualization and human evaluation.

**Comparison with the State of the Art:** It is essential to highlight that static fixation heatmaps are generated based on predicted fixation coordinates and fixation duration for each case. The intensity around each fixation coordinate is adjusted by scaling it with the fixation duration. For Table 2, we set the intensity spread around each fixation coordinate to 50. However, we also evaluated performance across all pixel spread levels and provided the comparison in Figure 3.

As shown in Table 2, when trained and tested on the same dataset (same radiologist), MedGaze shows significant improvements over Gazeformer [3]. Specifically, for the EGD-CXR dataset, MedGaze achieves a mIoU of  0.41 [95% CI 0.40,0.42 ], mCC of 0.50 [95% CI 0.48,0.51 ], mMM of 0.80 [95% CI 0.79,0.81 ], and mD-MM of 0.50 [95% CI 0.46,0.52 ], compared to Gazeformer's 0.27 [95% CI 0.26,0.28 ], 0.37 [95% CI 0.36,0.41 ], 0.71 [95% CI 0.70,0.71 ], and 0.06 [95% CI 0.048, 0.0839], respectively. On the REFLACX dataset, MedGaze achieves a mIoU of 0.45 [95% CI 0.44,0.46 ], mCC of 0.53 [95% CI 0.50,0.55 ], mMM of 0.84 [95% CI 0.83,0.85 ], and mD-MM of 0.66 [95% CI 0.65,0.68 ], while Gazeformer achieves 0.30 [95% CI 0.29,0.30 ], 0.40 [95% CI 0.38,0.42 ], 0.76 [95% CI 0.75,0.77 ], and 0.29 [95% CI 0.27,0.33 ], respectively. This substantial performance gain highlights MedGaze's superior ability to predict radiologists' scanpaths and fixation durations accurately.

Additionally, we assess performance based on dataset transferability to understand how well the model generalizes across different datasets. Since the EGD-CXR and REFLACX datasets are recorded by different radiologists, it is crucial to comprehend how well the model identifies abnormal regions corresponding to text, rather than solely overfitting to a specific dataset. When trained on EGD-CXR and tested on REFLACX, MedGaze achieves a mIoU of 0.39 [95% CI 0.38, 0.40] and an mCC of 0.42 [95% CI 0.40, 0.43], outperforming Gazeformer, which scores 0.26 [95% CI 0.25, 0.27] and 0.33 [95% CI 0.31, 0.34], respectively. Conversely, when trained on REFLACX and tested on EGD-CXR, MedGaze scores 0.41 (95% CI 0.40, 0.43) for mIoU, 0.50 (95% CI 0.47, 0.51) for mCC surpassing Gazeformer's 0.28 (95% CI 0.27, 0.29), 0.38 (95% CI 0.36, 0.41) respectively.

We also trained and tested our proposed model on the combined REFLACX+EGD-CXR dataset to evaluate whether a larger dataset would enhance scanpath predict [95% CI 0.48, 0.51], mMM of 0.85 [95% CI 0.84, 0.86], and mD-MM of 0.73 [95% CI 0.72, 0.74], significantly surpassing Gazeformer's scores of 0.30 [95% CI 0.29, 0.31], 0.42 [95% CI 0.40, 0.43], 0.78 [95% CI 0.77, 0.79], and 0.43 [95% CI 0.41, 0.45], respectively.

In Figure 2A, it is evident that MedGaze outperforms Gazeformer across all spread levels when both models are trained and tested on the same radiologist's data. In Figure 2B, MedGaze also surpasses Gazeformer across all spread levels when trained and tested on different radiologists' eye gaze datasets. Notably, the blue curve, representing MedGaze trained on a combination of both EGD-CXR and REFLACX, consistently outperforms all other curves. The orange curve, slightly below, represents MedGaze trained on EGD-CXR and tested on REFLACX. The difference between these curves (EGD_REF_MedGaze and EGD+Ref MedGaze) is more pronounced than the difference between the curves representing EGD+REF_Gazeformer and REF_EGD_Gazeformer. This indicates that MedGaze exhibits greater effectiveness and generalization when data augmentation is performed.

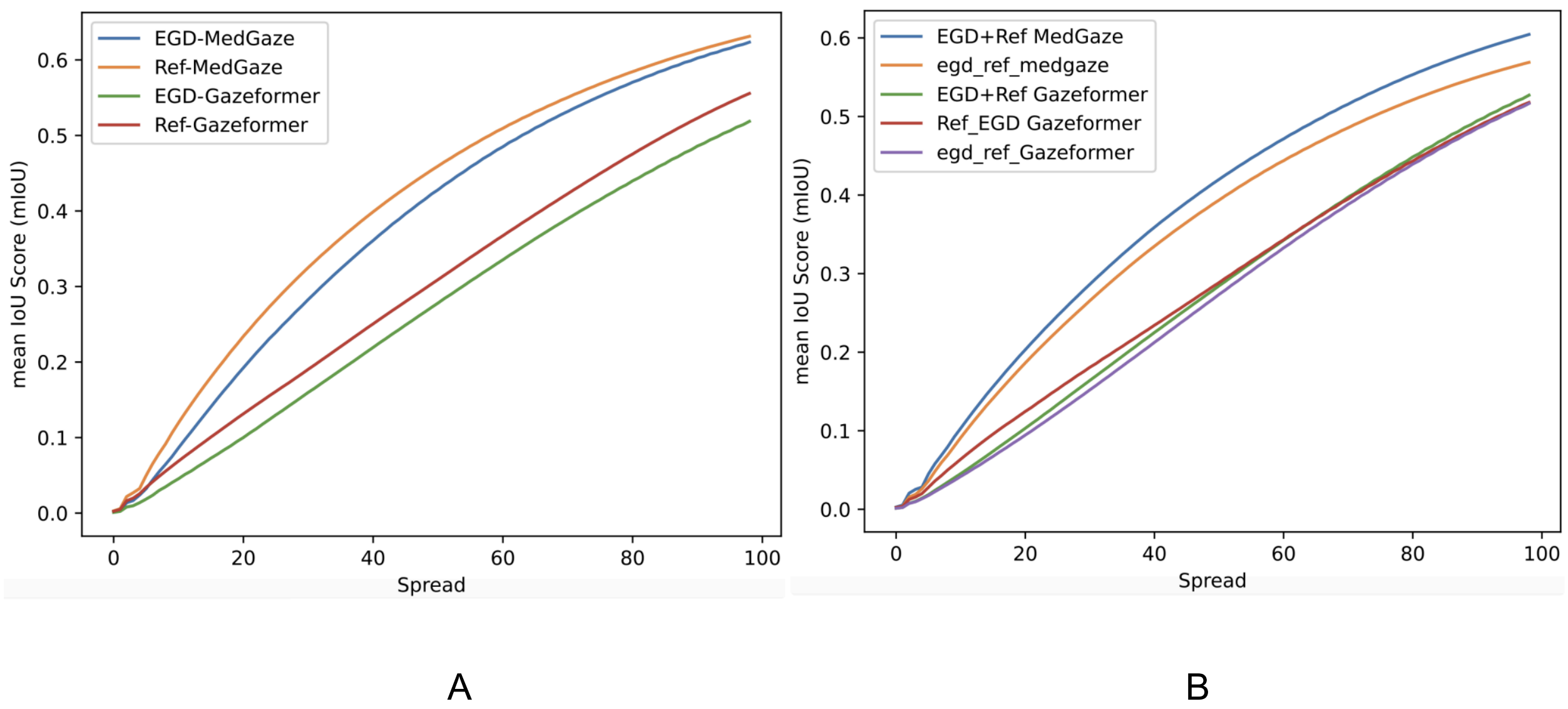

**Fig. 2: Figure represents two Subplots ( A, B ): Subplot A depicts the IoU comparison between MedGaze and the Gazeformer across different spread levels for models trained and tested on the same dataset. Subplot B illustrates the IoU comparison between MedGaze and the Gazeformer across different spread levels for models trained and tested on different datasets.**

## Predicted Scanpaths Visualization:

In Figure 3, visualized samples are derived from the test set from EGD-CXR data. As mentioned earlier, we set the maximum scanpath length to 50. This is an order of magnitude larger than Gazeformer's scanpath length for natural images. However, it is noteworthy that sometimes the ground truth scanpaths may exceed this predicted length, particularly when doctors continue examining the CXR image post-diagnosis. Nevertheless, our model can capture their post-interpretation analysis and incorporate it within the 50 scanpaths limit. Our model primarily focuses on predicting scanpaths corresponding to vital regions of interest rather than encompassing the entire image. Occasionally, the ground truth scanpaths may appear random and spread across all regions. For example, in the second row of Figure 3, where the radiologist mentions the 'right lower lung opacity is suspicious for pneumonia', our model accurately predicts fixation points around the right lower lung, indicating a higher opacity.

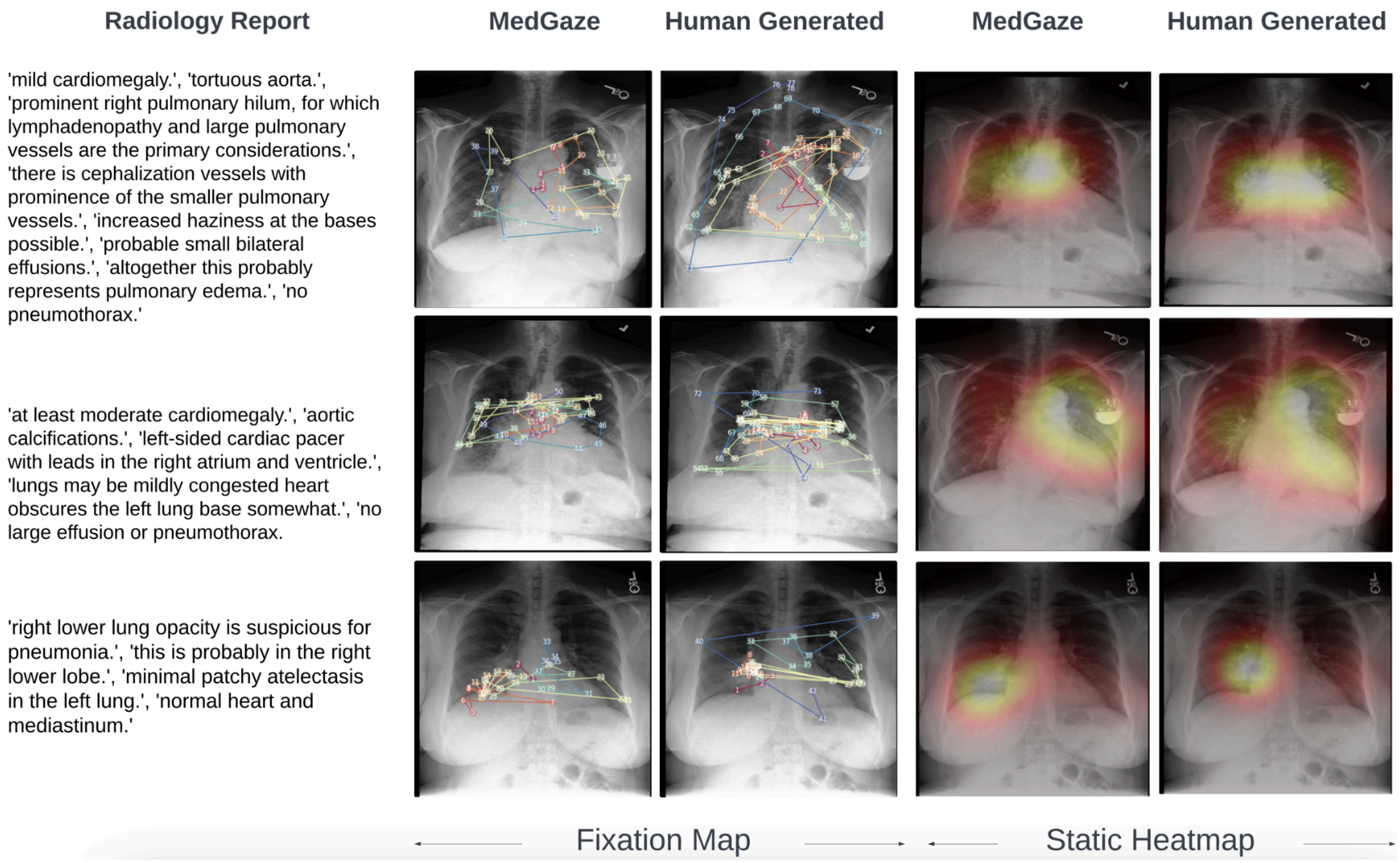

**Fig. 3: Illustration of predicted and ground truth scanpaths scaled with fixation duration alongside corresponding static fixation heatmaps. The first column displays the radiology report, while the second and third columns represent the predicted and ground truth fixation coordinates, respectively. In these columns, the red color arrow represents the start of the scan and the blue represents the end of the scan. The fourth and fifth columns show the predicted and ground truth static fixation heatmaps for the entire report, respectively.**

**Analyzing the Case Difficulty based on the fixation duration:**

Our investigation into case difficulty, inferred from fixation duration, reveals insightful findings regarding the model's comprehension of case difficulty. When trained and tested on the EGD-CXR dataset with recordings from a single experienced radiologist, the Pearson correlation coefficient (CC) between the radiologist's time duration and the model's predicted time duration was 0.54 (p=0), as illustrated in Figure 4, Column 1. Conversely, the REFLACX dataset, with recordings from five radiologists of varying experience levels, yielded a lower CC of 0.36 (p=0). To further assess case difficulty, we ranked cases based on total predicted fixation durations, with longer durations indicating higher difficulty (longer visual attention), and plotted these ranks against the ground truth. Furthermore, for the EGD-CXR dataset, a significant positive correlation between predicted and ground ranks was evident, with a Spearman rank correlation coefficient of 0.64 (p=0), depicted in Figure 4, Column 2. In the REFLACX dataset, the analysis showed a lower Spearman rank correlation coefficient of 0.36 (p=0), likely due to the dataset's inherent noise from multiple radiologists with varying expertise levels.

Figure 5 illustrates cases from the EGD-CXR test set positioned at both extremes (lowest and highest) of the distribution shown in Figure 4, Column 2, which represents the rank correlation. The cases ranked highest, indicating the most difficult scenarios, typically feature multiple abnormalities, thus enhancing their difficulty. In contrast, cases ranked lowest, denoting the simplest scenarios, frequently involve no abnormalities or represent normal cases.

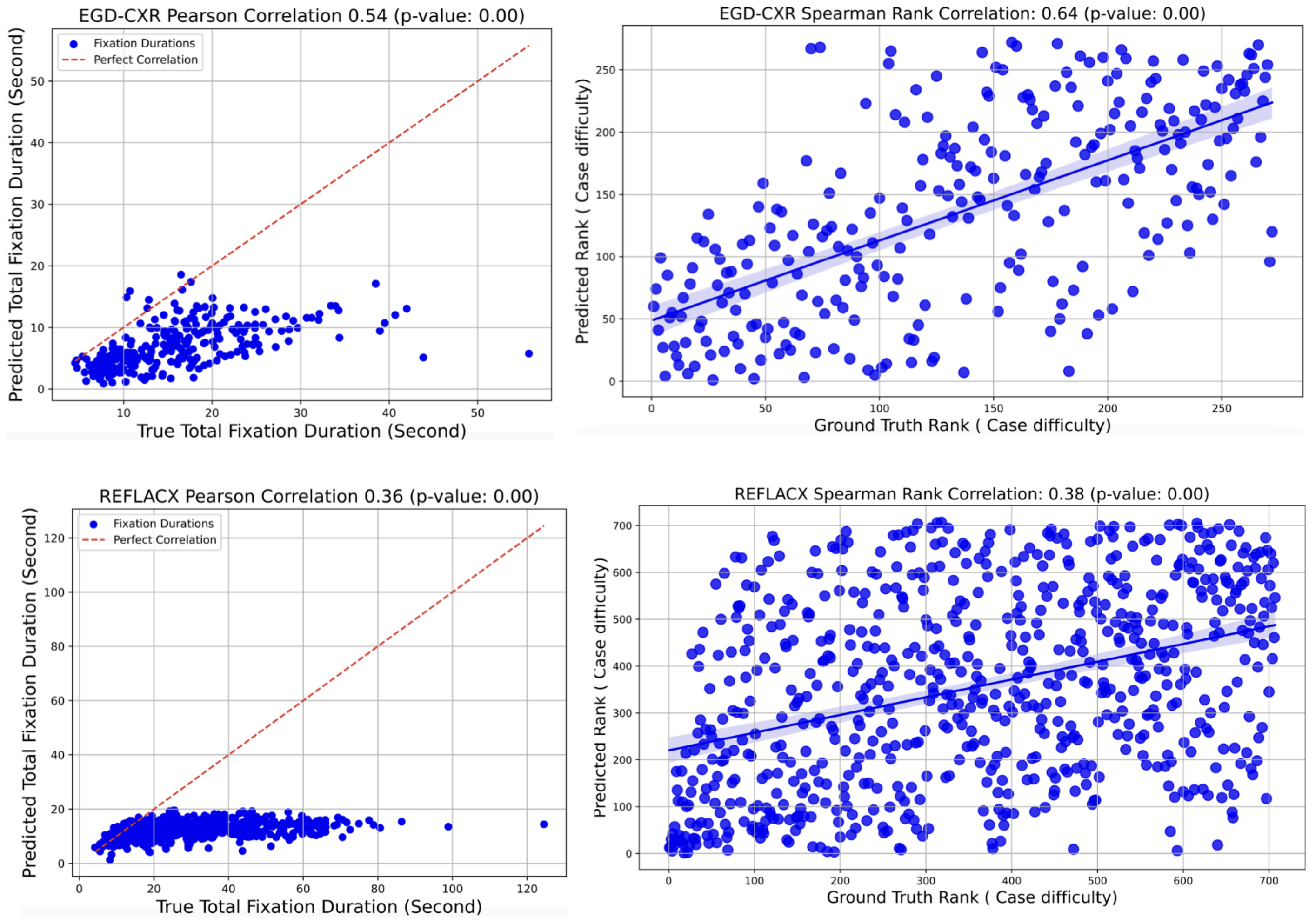

A B

**Fig. 4: Case Difficulty analysis using the Correlation coefficients. The figure comprises two columns (A and B ). Column A represents the Pearson Correlation Coefficient between true and predicted total fixation duration for the EGD-CXR and REFLACX test**

**sets . Column B represents the Spearman Rank Correlation between the true and predicted case difficulty ranks on the EGD-CXR and REFLACX test set.**

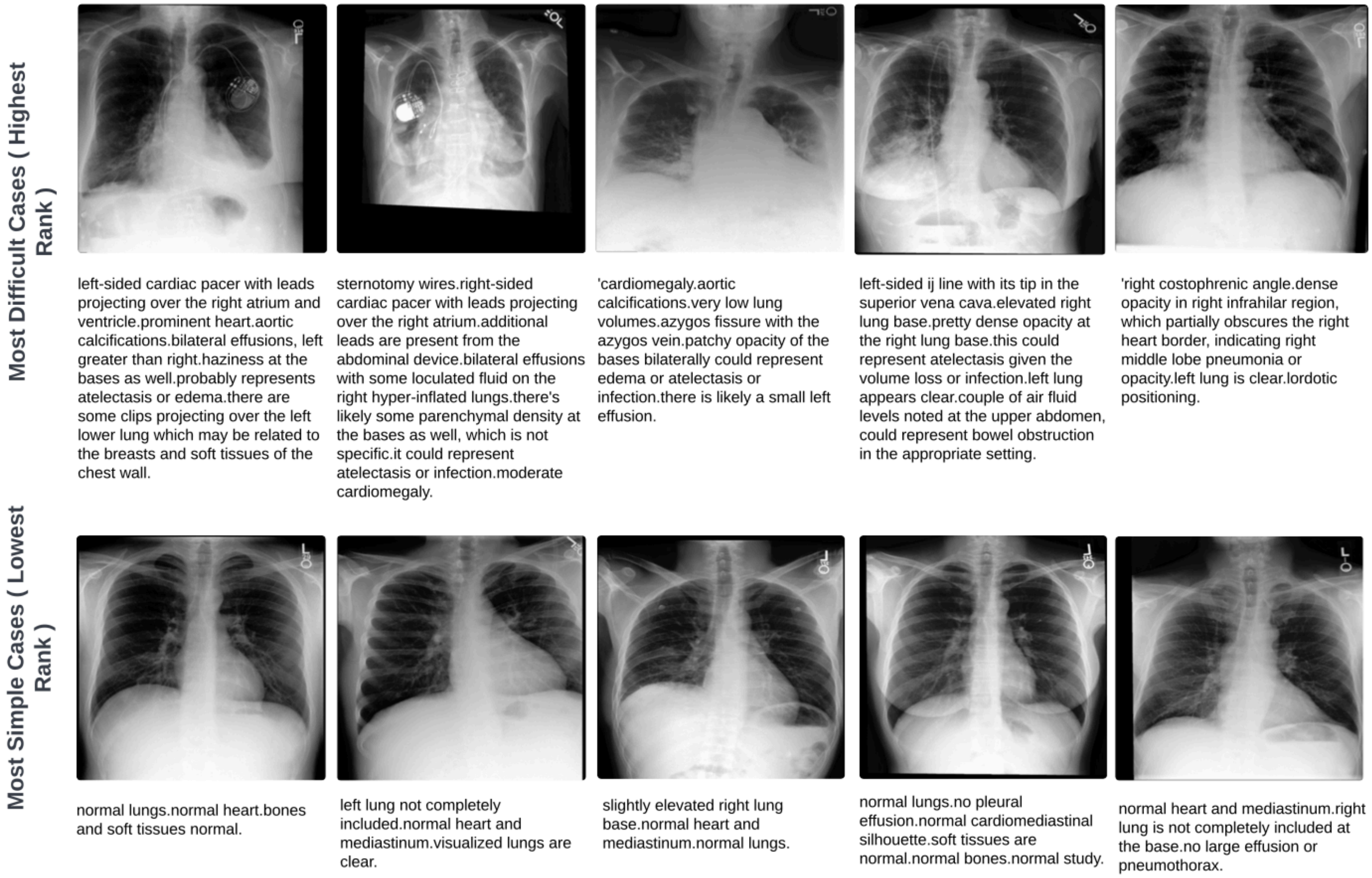

**Figure 5: Examples from the EGD-CXR dataset test set, illustrating the most challenging and simplest cases based on their rank correlations from Figure 4, Column 2. The first two rows show chest X-ray (CXR) images and corresponding radiology reports for cases ranked highest (most difficult), characterized by multiple abnormalities, and requiring longer attention. The third and fourth rows display CXR images and reports for cases ranked lowest (simplest), typically showing no abnormalities or normal findings.**

**Human Radiologist Evaluation:** We conducted a randomized control study where a board-certified radiologist was asked to score scanpaths without the knowledge of whether they are from human radiologists or MedGaze. The results, as shown in Table 3, indicate MedGaze's strong alignment with human gaze patterns. For identifying machine-generated versus human gaze patterns, MedGaze's predictions were rated as human-like in 13 out of 20 instances by a radiologist, compared to 19 for the ground truth, demonstrating a high degree of human-likeness. In terms of comprehensiveness, MedGaze showed robust coverage of important regions, with 8 predictions scoring a 4 (61-80% coverage) and 10 achieving a 5 (81-100% coverage), closely matching the ground truth, which had 8 and 12 predictions in these categories, respectively. Furthermore, MedGaze exhibited minimal redundancy, with most scores at 1 or 2, indicating efficient coverage with less overlap compared to human patterns, which had more instances of moderate redundancy. Overall, MedGaze effectively mimics the human gaze while maintaining efficiency and thorough coverage of significant regions. We also provide the randomly selected 40 video and expert radiologist evaluations in the source code repository.

| Criteria | Rating Scale | Prediction | Ground truth |
|---|---|---|---|
| Identifying Machine-Generated vs. Human Gaze Patterns | 0: (Machine-Generated) | 7 | 1 |
| | 1: (Human-Like) | 13 | 19 |
| Comprehensive Scores: Coverage of Important Regions | 1: (00-20%) Very little coverage | 0 | 0 |
| | 2: (21-40%) Some regions covered | 0 | 0 |
| | 3: (41-60%) Fair amount of coverage | 2 | 0 |
| | 4: (61-80%) Most regions covered | 8 | 8 |
| | 5: (81-100%) All regions covered | 10 | 12 |
| Redundancy Score: Coverage of Redundant Regions | 1: Minimal redundancy | 9 | 5 |
| | 2: Some minor redundancy | 7 | 11 |
| | 3: Moderate redundancy | 3 | 4 |
| | 4: Significant redundancy | 1 | 0 |
| | 5: High redundancy and inefficiency | 0 | 0 |

**Table 3:** Human Evaluation of MedGaze Predictions Compared to Human-Generated Scanpaths on CXR Images Across Defined Metrics

## Discussion:

This study introduces MedGaze, a novel system designed to model the complex cognitive processes of radiologists when interpreting chest X-ray (CXR) images. MedGaze employs a two-stage training strategy: Vision-Language Representation Learning and Vision Cognitive Learning. Initially, MedGaze is pre-trained on the publicly available MIMIC dataset to learn medically relevant multimodal features. Subsequently, the pre-trained MedGaze undergoes end-to-end training with the EGD-CXR and REFLACX datasets, aiming to predict scanpaths over CXR images. Our system is thoroughly evaluated using statistical metrics and human evaluation.

The table presents a performance comparison between MedGaze and the state-of-the-art (SOTA) method Gazeformer across different train/test combinations. Notably, when trained and tested on the same datasets (either EGD-CXR or REFLACX), MedGaze consistently outperforms Gazeformer in all metrics: mean Intersection over Union (mIoU), mean Correlation Coefficient (mCC), mean Multimatch Metric (mMM), and mean Duration Multimatch Metric (mD-MM). For instance, on the REFLACX dataset, MedGaze achieves a mIoU of 0.45 [95% CI 0.44, 0.46], an mCC of 0.53 [95% CI 0.50, 0.55], an mMM of 0.84 [95% CI 0.83, 0.85], and an mD-MM of 0.66 [95% CI 0.65, 0.68], significantly higher than Gazeformer's 0.30 [95% CI 0.29, 0.30], 0.40 [95% CI 0.38, 0.42], 0.76 [95% CI 0.75, 0.77], and 0.29 [95% CI 0.27, 0.33], respectively. The substantial difference in mD-MM scores for both datasets, EGD-CXR (MedGaze: 0.50 [95% CI 0.46, 0.52] vs. Gazeformer: 0.06 [95% CI 0.048, 0.0839]) and REFLACX (MedGaze: 0.66 [95% CI 0.65, 0.68] vs. Gazeformer: 0.29 [95% CI 0.27, 0.33]), highlights MedGaze's superior ability to predict fixation duration, crucial for understanding case difficulty. This performance can be attributed to our two-stage training approach, which effectively captures the intricate visual attention patterns of radiologists.

Our results also highlight the impact of dataset size on model performance. Both MedGaze and Gazeformer exhibit enhanced performance when trained on the larger REFLACX dataset compared to the smaller EGD-CXR dataset. This discrepancy is particularly evident in the metrics, with MedGaze's performance on REFLACX (mMM of 0.84 [95% CI 0.83, 0.85]) surpassing that on EGD-CXR (mMM of 0.80 [95% CI 0.79, 0.81]). This finding underscores the importance of large, diverse training datasets in improving model accuracy and generalizability.

Another crucial aspect of our study is MedGaze's ability to generalize across different radiologists. When trained on one dataset and tested on another (e.g., trained on REFLACX and tested on EGD-CXR), MedGaze still demonstrates robust performance, albeit with a slight decrease compared to training and testing on the same dataset. For example, MedGaze's mIoU drops from 0.45 (95% CI 0.44, 0.46) when trained and tested on REFLACX to 0.41 [95% CI 0.40, 0.43] when trained on REFLACX and tested on EGD-CXR.

To further validate MedGaze's effectiveness, we created a larger dataset by combining the REFLACX and EGD-CXR datasets. MedGaze achieved a mIoU of 0.41 [95% CI 0.40, 0.42], an mCC of 0.49 [95% CI 0.48, 0.51], an mMM of 0.85 [95% CI 0.84, 0.86], and an mD-MM of 0.73 [95% CI 0.72, 0.74], significantly higher than Gazeformer's scores of 0.30 [95% CI 0.29, 0.31], 0.42 [95% CI 0.40, 0.43], 0.78 [95% CI 0.77, 0.79], and 0.43 [95% CI 0.41, 0.45], respectively. Although there is a slight decrease in performance when combining data from different radiologists compared to training and testing on the same radiologist's data, our model still showed good performance. This suggests that combining data from multiple radiologists acts as a regularizer, introducing noise into the training process and aiding in generalizing the model across multiple datasets.

Radiologists' interpretation of CXRs entails varying fixation durations, influenced by multiple factors such as the complexity of findings, the number of abnormalities present, and their level of expertise, etc. To evaluate a model's ability to grasp case difficulty, we analyzed the Pearson correlation coefficient between predicted and ground truth total fixation durations. When tested on the EGD-CXR dataset (collected on experienced radiologist) , the model demonstrated a significant positive correlation (CC=0.56, p=0), indicating its tendency to predict longer durations for challenging cases. Conversely, testing on the REFLACX dataset, which features recordings from radiologists with varying expertise levels, resulted in a lower correlation coefficient (CC=0.35, p=0), reflecting the dataset's noise due to differing levels of radiologist experience. This noise impacted the model's ability to accurately learn case difficulty, despite its capacity to discern important regions amidst noisy data.

Expanding our investigation, we ranked case difficulty based on total fixation duration, with longer fixation durations corresponding to more difficult cases, thus representing higher ranks. For the EGD-CXR dataset, a significant positive correlation between predicted and ground truth ranks was evident, with a Spearman rank correlation coefficient of 0.56 (p=0). The rank-order plot illustrated a discernible trend, demonstrating how predicted ranks align well with ground truth ranks, particularly for cases with higher fixation durations. Subsequently, our analysis extended to the REFLACX dataset. Despite a lower Spearman rank correlation coefficient of 0.36 (p=0), likely attributed to the dataset's inherent noise from multiple radiologists of varying expertise levels. Additionally, we plotted the most difficult (highest rank) and simplest cases. The most difficult cases typically exhibit multiple abnormalities requiring careful attention for accurate diagnosis, whereas the simplest cases often depict normal conditions without abnormalities. This ranking approach can effectively guide the development of training programs for novice radiologists by presenting cases in increasing order of difficulty. Beginning with straightforward cases( normal cases with no abnormality) allows beginners to grasp normal anatomy and basic abnormalities, progressing to more challenging cases with longer fixation durations to refine their skills in identifying subtle or atypical findings. Such structured training enhances diagnostic accuracy and confidence, equipping radiologists to effectively manage diverse clinical scenarios while fostering continuous professional development.

The evaluation of MedGaze using human-likeness and comprehensiveness criteria reveals insightful findings about its performance in predicting gaze patterns. MedGaze's predictions were rated as human-like in 13 out of 20 cases, compared to 19 out of 20 for the ground truth, indicating a high degree of accuracy in emulating human gaze behavior. In terms of comprehensiveness, MedGaze demonstrated strong coverage of important regions, with 8 predictions scoring a 4 (61-80% coverage) and 10 predictions achieving a perfect score of 5 (81-100% coverage). This performance is comparable to the ground truth, where 8 and 12 predictions scored 4 and 5, respectively. However, the redundancy scores suggest that MedGaze predictions are less redundant than human gaze patterns, with a majority of its scores falling between 1 and 2 (minimal to some minor redundancy), while human patterns had more instances of moderate redundancy. This indicates that MedGaze not only effectively identifies crucial regions but also does so more efficiently, avoiding unnecessary fixation on redundant areas. Overall, the results underscore MedGaze's ability to closely mimic human gaze patterns while enhancing efficiency in gaze prediction.

Despite the promising results, several limitations must be acknowledged. The datasets used (REFLACX and EGD-CXR) are limited in size and diversity, potentially affecting the model's generalizability. The eye-tracking data, derived from a small number of radiologists, may not fully capture the complexity of human visual behavior. Additionally, MedGaze currently focuses solely on chest X-rays, and its applicability to other medical imaging modalities remains to be explored. Furthermore, the computational cost and complexity associated with large multimodal models could limit real-time clinical deployment.

In conclusion, MedGaze represents a significant advancement in predicting scanpaths on medical images, particularly chest X-rays (CXR). Through a two-stage training process leveraging large publicly available datasets, MedGaze accurately models the cognitive processes of radiologists to predict fixation coordinates and durations. Our system demonstrates superior performance compared to the state-of-the-art Gazeformer, as evidenced by higher IoU, CC, and Multimatch scores across different datasets. This improvement is consistent whether MedGaze is trained and tested on data from the same radiologist or across multiple radiologists, indicating robust generalizability. Furthermore, human evaluations affirm that MedGaze's predicted scanpaths closely resemble expert search patterns, with higher comprehensiveness and lower redundancy. These findings underscore the potential of MedGaze to enhance diagnostic accuracy, improve training programs for novice radiologists, and optimize clinical workflows, ultimately contributing to better patient outcomes. Future work will focus on expanding dataset diversity, exploring applicability to other imaging modalities, and optimizing the model for real-time clinical use.